\documentclass[twocolumn,showpacs,amsmath,aps,pra]{revtex4}

 \usepackage{graphicx}
 \usepackage{bm}

\begin{document}

 \title{Vacuum force on an atom in a magnetodielectric cavity}
 \author{M. S. Toma\v s}
 \email{tomas@thphys.irb.hr}
 \affiliation{Rudjer Bo\v skovi\' c Institute, P. O. B. 180,
 10002 Zagreb, Croatia}
 \date{\today}

 \begin{abstract}
We demonstrate that, according to a recently suggested
Lorentz-force approach to the Casimir effect, the vacuum force on
an atom embedded in a material cavity differs substantially from
the force on an atom of the cavity medium. The force on an
embedded atom is of the familiar (van der Waals and
Casimir-Polder) type, however, more strongly modified by the
cavity medium than usually considered. The force on an atom of the
cavity medium is of the medium-assisted force type with rather
unusual properties, as demonstrated very recently [M. S. Toma\v s,
Phys. Rev. A {\bf 71}, 060101(R) (2005)]. This implies similar
properties of the vacuum force between two atoms in a medium.
\end{abstract}
 \pacs{12.20.Ds, 42.50.Nn, 42.60.Da}
 \preprint{IRB-TH-3/05}
 \maketitle

It is well known that a neutral atom in the vicinity of a body
(mirror) experiences the van der Waals force \cite{Len} and, at
larger distances, its retarded counterpart, the Casimir-Polder
\cite{CP} force. This, commonly called the van der Waals force,
was considered theoretically numerous times using various methods
and for increasingly more complex systems
\cite{Lif,Boy,Schw,Zhou,Mil,Buh}. Apparently, the van der Waals
force is also quantitatively well supported experimentally
\cite{Ori,San,Suk,Lan,Shi,Dru,Lin,Pas}.

To account for the force on the cavity medium, which is absent in
the traditional approaches  to the Casimir effect \cite{Cas} in
material cavities \cite{Schw,Zhou,Mil,LiPi,Abr,Tom02}, Raabe and
Welsch \cite{Raa04} recently suggested a Lorentz-force approach to
the Casimir effect (see also Ref. \cite{Obu}). As an application
of this approach, Raabe and Welsch derived a formula for the force
on a magnetodielectric slab in a magnetodielectric planar cavity,
as depicted in Fig. 1. Applying their formula to a thin slab, in
this work we derive a general expression for the force on an
(electrically and magnetically polarizable) atom in a
magnetodielectric planar cavity. We demonstrate that, according to
this result, the force on an atom is substantially different,
depending on whether the atom is embedded in the cavity medium or
whether it is a constituent of the cavity medium. The force on the
embedded atom behaves in the familiar, although more strongly
modified by the cavity medium than found previously, way with the
atom-mirror distance and the electric/magnetic properties of the
atom and the mirror. Contrary to this, the force on an atom of the
cavity medium is a recently introduced medium-assisted force
\cite{Tom05}, with very unusual properties. We derive and discuss
a number of basic formulas concerning the atom-mirror force in
these two cases and establish a connection of these results with
their counterparts obtained through a traditional approach. We
also address shortly the implications of the obtained results on
the properties of the atom-atom force in a medium.

\begin{figure}[htb]
\label{sys}
 \begin{center}
 \resizebox{8cm}{!}{\includegraphics{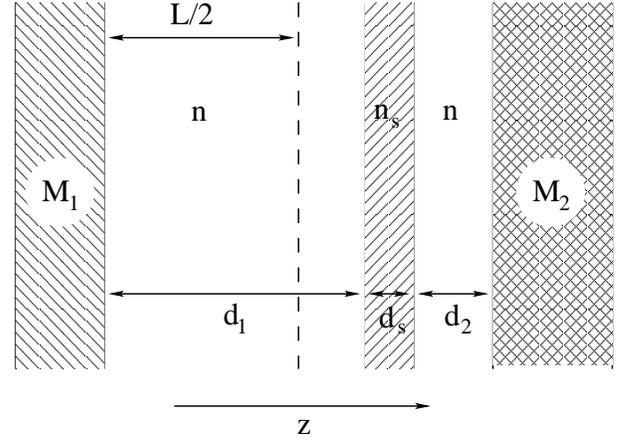}}
 \end{center}
 \caption{A slab in a planar cavity shown schematically. The
 (complex) refraction index of the slab is
 $n_s(\omega)=\sqrt{\varepsilon_s(\omega)\mu_s(\omega)}$ and that
 of the cavity $n(\omega)=\sqrt{\varepsilon(\omega)\mu(\omega)}$.
 The cavity walls are described by their reflection coefficients
 $r^q_1(\omega,k)$ and $r^q_2(\omega,k)$, with $k$ being the
 in-plane wave vector of a wave. The arrow indicates the direction
 of the force on the slab.}
\end{figure}

The Raabe and Welsch formula \cite{Raa04} for the force on the
slab in the configuration of Fig. 1 can be written as \cite{Tom05}
\begin{equation}
\label{f} f(d_1,d_2)=f^{(1)}(d_1,d_2)+f^{(2)}(d_1,d_2),
\end{equation}

\[f^{(1)}(d_1,d_2)=\frac{\hbar}{2\pi^2}\int_0^\infty
d\xi \int^\infty_0dkk\kappa\times\]
\begin{equation}
\label{f1}
\sum_{q=p,s}\left(\frac{1}{\varepsilon}\delta_{qp}+\mu\delta_{qs}\right)
r^q\frac{r^q_2e^{-2\kappa d_{2}}-r^q_1e^{-2\kappa d_{1}}}{N^q},
\end{equation}

\[f^{(2)}(d_1,d_2)=\frac{\hbar}{8\pi^2c^2}\int_0^\infty
 d\xi\xi^2\mu(n^2-1)\int^\infty_0\frac{dkk}{\kappa}\times\]
\begin{equation}
\label{f2}
 \sum_{q=p,s}[(1+r^q)^2-{t^q}^2]\Delta_q
\frac{r^q_2e^{-2\kappa d_{2}}-r^q_1e^{-2\kappa d_{1}}}{N^q},
 \end{equation}
where
\begin{eqnarray}
\label{N}
N^q&=&1-r^q(r^q_1e^{-2\kappa d_{1}}+r^q_2e^{-2\kappa
d_2})\nonumber\\
&+&({r^q}^2-{t^q}^2)r^q_1r^q_2e^{-2\kappa (d_1+d_2)}
\end{eqnarray}
and $\Delta_q=\delta_{qp}-\delta_{qs}$. Here
\begin{equation}
 \kappa(\xi,k)=\sqrt{n^2(i\xi)\frac{\xi^2}{c^2}+k^2}
\end{equation}
is the perpendicular wave vector in the cavity at the imaginary
frequency, $r^q$ and $t^q$ are Fresnel coefficients for the
(whole) slab given by
\begin{eqnarray}
\label{rt} r^q(i\xi,k)&=&\rho^q\frac{1-e^{-2\kappa_s
d_s}}{1-{\rho^q}^2e^{-2\kappa_s d_s}},\nonumber\\
t^q(i\xi,k)&=&\frac{(1-{\rho^q}^2)e^{-\kappa_s d_s}}
{1-{\rho^q}^2e^{-2\kappa_sd_s}},
\end{eqnarray}
where
\begin{equation}
\rho^q(i\xi,k)=\frac{\kappa-\gamma^q\kappa_s}
{\kappa+\gamma^q\kappa_s},\;\;\;
\gamma^p=\frac{\varepsilon}{\varepsilon_s},\;\;\;
\gamma^s=\frac{\mu}{\mu_s}, \label{rho}
\end{equation}
are the single-interface medium-slab Fresnel reflection
coefficients.

Equation (\ref{f1}) differs from the formula for the Casimir force
obtained through the Minkowski tensor calculation \cite{Tom02} in
the presence of the factors $1/\varepsilon$ and $\mu$ which
multiply the contributions coming from TM- and TE-polarized waves,
respectively. Effectively, these factors  diminish the force, so
that $f^{(1)}$ may be regarded as a medium-screened force. The
force $f^{(2)}$ owes its appearance to the cavity medium, note
that it vanishes when $n=1$, and can therefore be regarded as a
medium-assisted force. Note that, since the factor
$(1+r^q)^2-{t^q}^2$ is always positive, the sign of each term in
Eq. (\ref{f2}) depends only on whether the corresponding mirror
$i$ is dominantly conducting ($\Delta_q r^q_i>0$) or whether it is
dominantly permeable ($\Delta_q r^q_i<0$).


The force on an atom in a material cavity can be obtained from the
above equations by assuming that the slab consists of a thin,
$d_s\Omega/c\ll 1$, layer \cite{cut} of the cavity medium with a
small number of foreign atoms embedded in it. Then, from Eqs.
(\ref{rt}) and (\ref{rho}) we find that to the first order in
$\kappa_sd_s\sim\Omega d_s/c$
\begin{equation}
\label{ths}
 r^q(i\xi,k)\simeq 2\rho^q\kappa_s d_s,\;\;\;\;
[(1+r^q)^2-{t^q}^2](i\xi,k)\simeq 2\frac{\kappa d_s}{\gamma^q}.
\end{equation}
Also, we have
\begin{equation}
\varepsilon_s(i\xi)=\varepsilon(i\xi)+4\pi N\alpha_e(i\xi),\;\;\;
\mu_s(i\xi)=\mu(i\xi)+4\pi N\alpha_m(i\xi),
\end{equation}
where $N$ is the atomic number density and $\alpha_{e(m)}$ the
electric (magnetic) polarizability of an atom. Accordingly, for
small $N\alpha_{e(m)}$
\begin{equation}
\kappa_s\simeq \kappa\left[1+2\pi
N(\alpha_e\mu+\alpha_m\varepsilon)
\frac{\xi^2}{\kappa^2c^2}\right]
\end{equation}
so that [Eq. (\ref{rho})]
\begin{equation}
\rho^p\simeq \frac{2\pi N}{\varepsilon}
\left[\alpha_e-(\alpha_e\mu+\alpha_m\varepsilon)\frac{\varepsilon
\xi^2}{2\kappa^2c^2}\right]
\end{equation}
and $\rho^s=\rho^p[\varepsilon\leftrightarrow\mu,
\alpha_e\leftrightarrow\alpha_m]$.

With the above approximations inserted into Eqs.
(\ref{f})-(\ref{N}), the force on the layer can be, to the first
order in $d_s$, written as
\begin{equation}
f(d_1,d_2)=f_M(d_1,d_2)+Nd_s f_a(d_1,d_2),
\end{equation}
where
\begin{eqnarray}
f_M(d_1,d_2)&=&\frac{\hbar d_s}{4\pi^2 c^2}\int_0^\infty
 d\xi\xi^2\mu(n^2-1)\int^\infty_0dkk\times\nonumber\\
 &&\left[{\cal R}^p(i\xi,k)-{\cal R}^s(i\xi,k)\right],
 \label{fm}
 \end{eqnarray}
is the force on the medium ($M$) layer without the embedded atoms
\cite{Tom05}, and
\begin{eqnarray}
\label{fa}
f_a(d_1,d_2)&=&\frac{\hbar}{\pi c^2}\int_0^\infty
d\xi\xi^2
\int^\infty_0dkk\times\\
&&\left\{\frac{1}{\varepsilon}\left[\alpha_e
\left(2\frac{\kappa^2c^2}{\varepsilon\xi^2}-\mu\right)
-\alpha_m\varepsilon\right]{\cal R}^p(i\xi,k)\right.
\nonumber\\
&&+\mu\left[\alpha_m
\left(2\frac{\kappa^2c^2}{\mu\xi^2}-\varepsilon\right)
-\alpha_e\mu\right]{\cal R}^s(i\xi,k)
\nonumber\\
&&+\left.\mu(n^2-1)\left[\alpha_e{\cal R}^p(i\xi,k)-\alpha_m{\cal
R}^s(i\xi,k)\right]\right\}, \nonumber
\end{eqnarray}
is the force on an embedded atom. In these equations,
 \begin{equation}
{\cal R}^q(i\xi,k)= \frac{r^q_2e^{-2\kappa d_2}-r^q_1e^{-2\kappa
d_1}} {1-r^q_1r^q_2e^{-2\kappa (d_1+d_2)}}.
 \end{equation}

Similarly as in Eqs. (\ref{f})-(\ref{f2}), the first two terms in
Eq. (\ref{fa}) describe a medium-screened force $f^{(1)}_a$ and
the last one a medium-assisted force $f^{(2)}_a$ on the atom.
Accordingly, the force on an atom in the Minkowski stress-tensor
approach, $f^{(M)}_a$, is obtained from the above result for
$f^{(1)}_a$ by removing the factors $1/\varepsilon$ and $\mu$ from
$p$ and $s$ contributions to the integrand, respectively. We
therefore have
\begin{eqnarray}
 \label{faM}
f^{(M)}_a(d_1,d_2)&=&\frac{\hbar}{\pi c^2}\int_0^\infty d\xi\xi^2
\int^\infty_0dkk\times\\
&&\left\{\left[\alpha_e
\left(2\frac{\kappa^2c^2}{\varepsilon\xi^2}-\mu\right)
-\alpha_m\varepsilon\right]{\cal R}^p(i\xi,k)\right.
\nonumber\\
&&+\left.\left[\alpha_m
\left(2\frac{\kappa^2c^2}{\mu\xi^2}-\varepsilon\right)
-\alpha_e\mu\right]{\cal R}^s(i\xi,k)\right\}, \nonumber
\end{eqnarray}
which coincides with the result obtained by Zhou and Spruch (using
the surface mode summation method) \cite{Zhou} but it is
generalized by accounting for the magnetic properties of the
system. Of course, both Eq. (\ref{fa}) and Eq. (\ref{faM}) give
the same result in the case of an empty ($n=1$) cavity.

Assuming, for simplicity, a semi-infinite cavity obtained by
removing, say, mirror 1 ($r^q_1=0$), we have
\begin{equation}
{\cal R}^q(i\xi,k)= R^q(i\xi,k)e^{-2\kappa z},
 \end{equation}
where $R^q\equiv r^q_2$ and $z\equiv d_2$ is the atom-mirror
distance. Then, owing to the above exponential factor, for small,
$z\ll c/\Omega$, atom-mirror distances \cite{cut} the major
contribution to the integral in Eq. (\ref{fa}) comes from large
$k$ values. Approximating the integrand with its nonretarded
($k\rightarrow\infty$) counterpart and making the substitution
$u=k/2d$, we find for the leading term of the atom-mirror force
\begin{eqnarray}
f_a(z)&=&\frac{\hbar}{8\pi z^4}\int_0^\infty
d\xi\int^\infty_0duu^3 e^{-u}\times\\
&&\left[\frac{\alpha_e}{\varepsilon^2} R^p_{\rm
nr}(i\xi,\frac{u}{2d})+ \alpha_m R^s_{\rm
nr}(i\xi,\frac{u}{2d})\right]
\label{fas}
\end{eqnarray}
where $R^q_{\rm nr}(i\xi,k)$ are reflection coefficients of the
mirror in the nonretarded approximation. These coefficients are
formally obtained from $R^q(i\xi,k)$ by letting $\kappa=k$ and
$\kappa_l=k$ for the perpendicular wave vectors in all layers of
the mirror. Specially, for a single-medium mirror with the
refraction index $n_m$, $R^q_{\rm nr}(i\xi,u/2z)$ are independent
of $u$ [see Eq. (\ref{rho}), with
$\{\varepsilon_s,\mu_s\}\rightarrow\{\varepsilon_m,\mu_m\}$].
Performing the elementary integration, in this classical
configuration we therefore find
\begin{equation}
f_a(z)=\frac{3\hbar}{4\pi z^4}\int_0^\infty d\xi
\left[\frac{\alpha_e}{\varepsilon^2}\frac{\varepsilon_m-\varepsilon}
{\varepsilon_m+\varepsilon}+ \alpha_m \frac{\mu_m-\mu}
{\mu_m+\mu}\right]. \label{fas2}
\end{equation}
This generalizes the familiar result for the van der Waals force
on an atom close to a medium to magnetodielectric systems. Note
the presence of the extra (screening) factors $1/\varepsilon$ and
$\mu$ in comparison with the corresponding traditionally obtained
formula \cite{Zhou}

To find $f_a(z)$ for large $z$, we make the standard substitution
$\kappa=n\xi p/c$ in Eq. (\ref{fa}) and, since $\xi\sim c/z$,
approximate the frequency-dependent quantities with their static
values (denoted by the subscript $0$). The integral over $\xi$ can
then be easily performed and we obtain
\begin{eqnarray}
\label{fal}
 f_a(z)&=&\frac{3\hbar c}{4\pi n^3_0z^5}
\int^\infty_1\frac{dp}{p^4}\times\\
&&\left\{\frac{1}{\varepsilon_0}\left[\alpha_{e0}\mu_0
(2p^2-1)-\alpha_{m0}\varepsilon_0\right]R^p(0,p)\right.
\nonumber\\
&&+\mu_0\left[\alpha_{m0}\varepsilon_0
(2p^2-1)-\alpha_{e0}\mu_0\right]R^s(0,p)
\nonumber\\
&&+\left.\mu_0(n^2_0-1)\left[\alpha_{e0}R^p(0,p)-\alpha_{m0}
R^s(0,p)\right]\right\}, \nonumber
\end{eqnarray}
where $R^q(i\xi,p)$ are obtained from $R^q(i\xi,k)$ by letting
$\kappa_l\rightarrow n(\xi/c)s_l$, with
$s_l=\sqrt{p^2-1+n^2_l/n^2}$ for all relevant layers. Thus, for
example, for a single-medium mirror we have
\begin{equation}
\label{rps} R^p(i\xi,p)=\frac{\varepsilon_m p-\varepsilon s_m}
{\varepsilon_m p+\varepsilon s_m},\;\;\; R^s(i\xi,p)=\frac{\mu_m
p-\mu s_m} {\mu_m p+\mu s_m}.
\end{equation}
To illustrate this result, we consider the case of an ideally
reflecting ($\varepsilon_m\rightarrow\infty$ or
$\mu_m\rightarrow\infty$) mirror. Letting $R^q=\pm\Delta_q$ (the
minus sign is for an infinitely permeable mirror) in Eq.
(\ref{fal}), we obtain 
\begin{eqnarray}
f^{\rm id}_a(z)&=&\pm\frac{\hbar c} {4\pi z^5n_0\varepsilon_0}
\left[\alpha_{e0}\left(\frac{5}{\varepsilon_0}+\mu_0+n^2_0-1
\right)\right.\nonumber\\
&-&\left.\alpha_{m0}\left(\frac{1}{\mu_0}+5\varepsilon_0-n^2_0+1
\right)\right], \label{faid}
\end{eqnarray}
whereas the "traditional" Eq. (\ref{faM}) in this case gives
\begin{equation}
f^{(M){\rm id}}_a(z)=\pm\frac{3\hbar c} {2\pi z^5n_0^3}
(\alpha_{e0}\mu_0-\alpha_{m0}\varepsilon_0) \label{fMaid}
\end{equation}
Of course, in the case of an empty ($\varepsilon=\mu=1$) cavity,
we recover from both these equations the Boyer generalization 
\cite{Boy} of the Casimir-Polder formula \cite{CP}.

The force on an atom of the cavity medium $f^M_a$ can be similarly
obtained from Eq. (\ref{fm}). Assuming a dilute medium, $n^2\simeq
1+4\pi N_M(\alpha^M_e+\alpha^M_m)$, we have $f^M_a=f_M/N_Md_s$,
where $N_M$ is the atomic number density in the cavity. This force 
is, to the first order in $\alpha^M_e$ and $\alpha^M_m$ and for a 
single-medium mirror, given by
\cite{Tom05}
\begin{eqnarray}
\label{fams} f^M_a(z)&=&\frac{\hbar}{4\pi c^2z^2}\int_0^\infty
d\xi\xi^2 (\alpha^M_e+\alpha^M_m)
\times \nonumber\\
&&\left[\frac{\varepsilon_m-1} {\varepsilon_m+1}-
\frac{\mu_m-1}{\mu_m+1}\right]
\end{eqnarray}
at small and by
\begin{eqnarray}
\label{faml} f^M_a(z)&=&\frac{3\hbar
c}{4\pi z^5}(\alpha^M_{e0}+\alpha^M_{m0})\times\nonumber\\
&&\int^\infty_1\frac{dp}{p^4} \left[\frac{\varepsilon_{m0}
p-s_{m0}} {\varepsilon_{m0} p+s_{m0}}-\frac{\mu_{m0} p-s_{m0}}
{\mu_{m0} p+s_{m0}}\right]
\end{eqnarray}
at large atom-mirror distances. Here $s_m=\sqrt{p^2-1+n^2_m}$, as
appropriate for a dilute cavity medium. Note that, besides
exhibiting very unusual behavior at small atom-mirror distances,
the sign of $f^M_a$ is completely insensitive to the
polarizability type of the atom. Since for an ideally reflecting
mirror the value of integral in Eq. (\ref{faml}) is $\pm 2/3$, we
see that $f^M_a$ is comparable with $f_a$ at large distances [c.f.
Eq. (\ref{faid})].

The above results imply similar properties of the force $f_{aa}$
between two atoms in a medium. This force can be found in the
usual way \cite{Lif,LiPi} by assuming a single-medium mirror
consisting of the cavity medium with a small number of, say, type
$B$ atoms embedded in it, so that $\varepsilon_m=\varepsilon+4\pi
N_B\alpha^B_e$ and $\mu_m=\mu+4\pi N_B\alpha^B_m$. With this
inserted in Eqs. (\ref{fas2}) and (\ref{fal}), we obtain small and
large distance behavior of the force $f^{AB}_{aa}$
[$\alpha_{e(m)}\equiv\alpha^A_{e(m)}$] between two atoms $A$ and
$B$ embedded in the medium and from Eqs. (\ref{fams}) and
(\ref{faml}) we obtain the corresponding behavior of the force
$f^{MB}_{aa}$ between an atom of the medium and an embedded atom .
Thus, for example, from Eq. (\ref{fas2}) we straightforwardly find
\cite{com2}
\begin{equation}
f^{AB}_{aa}(r)=\frac{18\hbar}{\pi r^7}\int_0^\infty d\xi
\left[\frac{1}{\varepsilon^3}\alpha^A_e\alpha^B_e+
\frac{1}{\mu}\alpha^A_m\alpha^B_m\right],
\end{equation}
which predicts stronger medium screening of the van der
Waals-London force than found earlier \cite{LiPi}, and from Eq.
(\ref{fams}) we obtain
\begin{equation}
f^{MB}_{aa}(r)=\frac{2\hbar}{\pi c^2r^5}\int_0^\infty d\xi\xi^2
(\alpha^M_e+\alpha^M_m)(\alpha^B_e-\alpha^B_m),
\end{equation}
which implies different properties of the interaction between an
atom of the medium and an embedded atom.

In summary, according to the Lorentz force approach to the Casimir
effect, the force on an atom embedded in a material cavity differs
substantially from the force on an atom of the cavity medium. For
embedded atoms, the force consists of a medium-screened and a
medium-assisted force. The results for the medium-screened force
differ from the corresponding traditionally obtained results in
the presence of the extra factors $1/\varepsilon$ and $\mu$
multiplying the contributions of the TM and TE polarized waves,
respectively. This, together with the appearance of the
medium-assisted force term, predicts a stronger dependence of the
atom-mirror force on the medium parameters than usually
considered. Accordingly, a number of the classical results for the
atom-mirror interaction in various systems are modified with
respect to this point. The force on the atoms of the cavity medium
is a very recently introduced medium-assisted force \cite{Tom05},
which behaves as the Coulomb force at small and as the
Casimir-Polder force at large atom-mirror distances. In addition,
contrary to the Casimir-Polder force \cite{Boy}, its sign is
insensitive to the polarizability type (electric or magnetic) of
the atom. Clearly, these properties of the atom-mirror force imply
similar properties of the atom-atom force in a medium.

{\it Note added}. After this work was completed, strong doubts
have been raised \cite{Pit} on the correctness of the stress
tensor (brackets denote the average with respect to fluctuations)
\begin{eqnarray}
 T_{ij}({\bf r})&=&\frac{1}{4\pi}\left<D_iE_j+H_iB_j-\frac{1}{2}
 \left({\bf D}\cdot{\bf E}+{\bf H}\cdot{\bf B}\right)\delta_{ij}\right>
 \nonumber\\
&-&\left<P_iE_j-M_iB_j-\frac{1}{2}
 \left({\bf P}\cdot{\bf E}-{\bf M}\cdot{\bf B}\right)\delta_{ij}\right>
 \label{T}
 \end{eqnarray}
employed by Raabe and Welsch \cite{Raa04}. As stressed by
Pitaevskii \cite{Pit}, the first (Minkowski) term here corresponds
to the effective stress tensor in a (fluid) medium which is in
mechanical equilibrium. If so, the above stress tensor is
incomplete since its second term, which gives rise to the force on
the medium, is not balanced. We note that such a conclusion is
also (implicitly) indicated by rather peculiar properties of the
van der Waals and Casimir forces implied by $T_{ij}$ given by Eq.
(\ref{T}), as we have demonstrated in this work as well as in Ref.
\cite{Tom05}.

This work was supported by the Ministry of Science and Technology
of the Republic of Croatia under contract No. 0098001.


\begin{thebibliography}{00}
\bibitem{cut} The integral over $\xi$ in Eq. (\ref{f1}) effecively
extends up to a frequency $\Omega$ beyond which either the slab or
the corresponding mirror becomes transparent, depending on which
of these two quantites is smaller. In Eq. (\ref{f2}), the
corresponding cutoff frequencies are given by the properties of
the mirror and the cavity medium. We can therefore take $\Omega$
of the mirror as the cutoff frequency common to both equations.
This introduces the characteristic length $c/\Omega$ by which all
other lengths (and thicknesses) in the system may be compared. For
a more rigorous consideration of this point, see, e.g., Ref.
\cite{Buh}.
\end{thebibliography}
\end{document}